\documentclass{article}
\usepackage{amsfonts}
\usepackage{amsmath}

\input{tcilatex}
\begin{document}

\begin{titlepage}
\title{\bf Mechanical Systems on an almost K\"{a}hler model of a Finsler Manifold}
\author{ Mehmet Tekkoyun \footnote{ Corresponding Author. tekkoyun@comu.edu.tr} \\
{\small Department of Econometrics, Faculty of Economics and Administration Sciences at Biga,}\\ {\small \c{C}anakkale Onsekiz Mart University}, {\small 17200 Biga/\c{C}anakkale, Turkey}\\
Oguzhan Celik\footnote{oguzhanefe07@hotmail.com} \\
 {\small Department of Mathematics, Faculty of Science and Art, Pamukkale University,}\\
 {\small Denizli, Turkey}}
\date{\today}
\maketitle

\begin{abstract}

In this study, we present a new analogue of Euler-Lagrange and
Hamilton equations on an almost K\"{a}hler model of a Finsler
manifold. Also, we give some corollories about the related
mechanical systems and equations.



\end{abstract}
\end{titlepage}

\section{Introduction}

In modern differential geometry, a suitable vector field defined on the
tangent and cotangent spaces (or manifolds) which are phase-spaces of
velocities and momentum of a given configuration space clarifies the
dynamics of Lagrangian and Hamiltonian. If $M$ is an $n$-dimensional
configuration manifold and $L(H):TM(T^{\ast }M)\rightarrow \mathbf{R}$%
\textbf{\ }is a regular Lagrangian (Hamiltonian) function, then there is a
unique vector field $\xi $ $(X)$ on tangent (cotangent) bundle $TM$ $%
(T^{\ast }M)$ such that dynamical formalisms and equations are given by%
\begin{equation*}
\begin{array}{cc}
\hline
\text{Dynamical Formalisms} & \text{Produced Dynamical Equations} \\ \hline
i_{\xi }\Phi _{L}=dE_{L} & \frac{\partial }{\partial t}(\frac{\partial L}{%
\partial \overset{.}{q}^{i}})-\frac{\partial L}{\partial q^{i}}=0 \\ \hline
i_{X}\Phi _{H}=dH & \frac{dp^{i}}{dt}=-\frac{\partial H}{\partial q^{i}},%
\mathit{\ }\frac{dq^{i}}{dt}=\frac{\partial H}{\partial p^{i}} \\ \hline
\end{array}%
\end{equation*}%
where $\Phi _{L}(\Phi _{H})$ indicates the (canonical) symplectic form. The
triples $(TM,\Phi _{L},\xi )$ and $(T^{\ast }M,\Phi _{H},Z_{H})$ are called 
\textit{Lagrangian system} and \textit{Hamiltonian system }on the tangent
bundle $TM$ and on the cotangent bundle $T^{\ast }M,$ respectively \cite%
{deleon1}$.$

Anyone notes that almost K\"{a}hler models related to Finsler geometry and
generalization were originally proposed by Matsumoto \cite{matsumoto}, then
the Finsler idea was generalized to Lagrange geometries by Kern \cite{kern},
and extended in Oproiu's works \cite{vasile, vasile1}$.$ But, in general,
almost K\"{a}hler - Finsler constructions and generalizations, with
applications to modern classical and quantum physics and mechanics where
performed in tenths of Vacaru's papers so that some of them are \cite%
{vacaru, vacaru1, vacaru2}.

Really, in \cite{miron, miron1, miron2} being the books of R. Miron, M.
Anastasiei and their other colleagues there are summarized a number of
results on almost K\"{a}hler - Lagrange - Finsler/ Hamilton - Cartan
geometries and generalizations with applications in mechanics.

Nevertheless, the key ideas were in Matsumoto - Kern papers and various
applications related to classical and quantum gravity were considered in
Vacaru's papers. There were also elaborated various Ricci flow, (non)
commutative and algebroid constructions related to Finsler models and
generalizations.

In \cite{udriste}, extrema of p-energy on constant curvature Finsler spaces
was investigated. In the studies \cite{tekkoyun1, tekkoyun2}, mechanical
systems of quaternionic K\"{a}hler and paracomplex geometry were analyzed
successfully. In this paper, the equations related to Finsler mechanical
systems have been presented and given some corollaries.

$\mathcal{F}(M)$, $\chi (M)$ and $\Lambda ^{1}(M)$ denote the set of
functions on manifold $M$, the set of vector fields on $M$ and the set of
1-forms on $M$, respectively.

\section{Finsler metric and Finsler Manifold}

In this section we recall some structures by given \cite{miron, vasile2}. A
Finsler manifold (or Finsler space) is a pair $F^{n}=(M,F(x,y))$ where $M$
is a differential manifold \ of real dimension $n$ and $F:TM\rightarrow 
%TCIMACRO{\U{211d} }%
%BeginExpansion
\mathbb{R}
%EndExpansion
$ a scalar function which supply :

$i)$ $F$ is a differentiable function on the manifold $\widetilde{TM}%
=TM\backslash \{0\}$ and $F$ is continuous on the null section of the
projection $\pi :TM\rightarrow M.$

$ii)$ $F$ is positive function.

$iii)$ $F$ is positively 1-homogeneous on the fibres of tangent bundle $TM$.

$iv)$ The Hessian of $F^{2}$ with elements%
\begin{equation}
g_{ij}(x,y)=\frac{1}{2}\frac{\partial ^{2}F^{2}}{\partial y^{i}\partial y^{j}%
}  \label{1.1}
\end{equation}

is positively defined on $\widetilde{TM}.$ Where $g_{ij}$ is a covariant
symmetric of 2 order distinguished tensor field (d-tensor field) defined on
the manifold $\widetilde{TM}.$

The function $F(x,y)$ is called fundamental function and the d-tensor field $%
g_{ij}$ is called fundamental (or metric) tensor of the Finsler space $%
F^{n}=(M,F(x,y)).$

A Finsler space $F^{n}=(M,F)$ can be thought as an almost K\"{a}hler space
on the manifold $\widetilde{TM}=TM\backslash \{0\}$ , called the geometrical
model of the Finsler space $F^{n}.$ If we consider the Cartan nonlinear
connection $N_{j}^{i}$ of the Finsler space $F^{n}=(M,F),$ then we can
respectively define almost complex structures $\mathbf{F}$ and $\  \mathbf{F}%
^{\ast }$ on $TM$ and $T^{\ast }M$ by: 
\begin{equation}
\mathbf{F}(\frac{\delta }{\delta x^{i}})=-\frac{\partial }{\partial y^{i}}\
,\  \  \mathbf{\ F}(\frac{\partial }{\partial y^{i}})=\frac{\delta }{\delta
x^{i}}  \label{1.2}
\end{equation}

\begin{equation}
\mathbf{F}^{\ast }\left( dx^{i}\text{ }\right) =-\delta y^{i}\ ,\  \  \ 
\mathbf{F}^{\ast }\left( \delta y^{i}\  \right) =dx^{i}  \label{1.3}
\end{equation}

Where the\ basis and dual basis are given as follows:

\begin{equation*}
\begin{array}{ccc}
\hline
TM & HTM & VTM \\ \hline
(\frac{\partial }{\partial x^{i}},\frac{\partial }{\partial y^{i}}) & \frac{%
\delta }{\delta x^{i}}=\frac{\partial }{\partial x^{i}}-N_{j}^{i}(x,y)\frac{%
\partial }{\partial y^{i}} & \frac{\partial }{\partial y^{i}}%
\end{array}%
\end{equation*}

\begin{equation*}
\begin{array}{ccc}
\hline
T^{\ast }M & HTM & VTM \\ \hline
(dx^{i},dy^{i}) & \delta y^{i}\ =dx^{i}+N_{j}^{i}(x,y)dy^{i} & dx^{i}%
\end{array}%
\end{equation*}

also $(\frac{\partial }{\partial x^{i}})^{H}=\frac{\delta }{\delta x^{i}};$ $%
(dy^{i})^{H}=\delta y^{i}$. It is easy to see that $\mathbf{F}$ is well
defined on $\widetilde{TM},$ $\mathbf{F}^{2}=-I$ \ and it is determined only
by the fundamental function $F$ of the Finsler space $F^{n}.$

Let $\left( dx^{i}\text{ ,}\delta y^{i}\right) $ be the dual basis of the
adapted basis $(\frac{\delta }{\delta x^{i}},\frac{\partial }{\partial y^{j}}%
).$ Then, the Sasaki-Matsumoto lift of the fundamental tensor $g_{ij}$ can
be introduced as follows:

\begin{equation}
G=g_{ij}dx^{i}\otimes dx^{j}+g_{ij}\delta y^{i}\otimes \delta y^{j}.
\label{1.4}
\end{equation}

Consequently, $G$ is a Riemann metric on $\widetilde{TM}$ determined only by
the fundamental function $F$ of the Finsler space $F^{n}$ and the horizontal
and vertical distributions are orthogonal with respect to it.

\textbf{Theorem: }

i) The pair $(G,\mathbf{F})$ is an almost Hermitian structure on $\widetilde{%
TM}.$

ii) The almost symplectic 2-form associated to the almost Hermitian
structure $(G,\mathbf{F})$ is

\begin{equation}
\theta =g_{ij}(x,y)\delta y^{i}\otimes dx^{j}.  \label{1.5}
\end{equation}

iii) The space $H^{2n}=(\widetilde{TM};G,\mathbf{F})$ is an almost K\"{a}%
hler space, constructed only by means of the fundamental function $F$ of the
Finsler space $F^{n}.$

\textbf{Definition: }The space $H^{2n}=(\widetilde{TM};G,\mathbf{F})$ is
called the almost K\"{a}hler model of the Finsler space $F^{n}.$

Remarking that the tensor field $\mathring{G}$ on $\widetilde{TM}$ given by

\begin{equation}
\mathring{G}=g_{ij}(x,y)dx^{i}\otimes dx^{j}+\frac{a^{2}}{\left \Vert
y\right \Vert ^{2}}g_{ij}(x,y)\delta y^{i}\otimes \delta y^{j},\text{ }%
\forall (x,y)\in \widetilde{TM}.  \label{1.6}
\end{equation}

is the homogeneous lift to $\widetilde{TM}$ of the fundamental tensor field $%
g_{ij}$ of a Finsler space $F^{n},$

where $a>0$ is a constant, imposed by applications (in order to preserve the
physical dimensions of the components of $\mathring{G}$) and where $%
\left
\Vert y\right \Vert
^{2}=g_{ij}(x,y)y^{i}y^{j}=y_{i}y^{i}=F^{2}(x,y)=2t$ with $%
y_{i}=g_{ij}(x,y)y^{j}=\frac{1}{2}\frac{\partial F^{2}}{\partial y^{i}} $
such that $t$ the value of energy density in $y\in TM.$

Let us prove that the almost complex structure $\mathbf{F}$, defined by (\ref%
{1.2}) does not preserve the property of homogeneity of the vector fields.
Indeed, it applies the 1-homogeneous vector fields $\frac{\delta }{\delta
x^{i}},(i=\overline{1,n})$ onto the 0-homogenous vector fields $\frac{%
\partial }{\partial y^{i}},(i=\overline{1,n}).$

We can eliminate this $\mathbf{F}$ by defining a new kind of almost complex
structure $\overset{\circ }{\mathbf{F}}:\chi (\widetilde{TM})\rightarrow
\chi (\widetilde{TM}),$ setting : 
\begin{equation}
\overset{\circ }{\mathbf{F}}(\frac{\delta }{\delta x^{i}})=-\frac{\left
\Vert y\right \Vert }{a}\frac{\partial }{\partial y^{i}}\ ,\  \  \mathbf{\ }%
\overset{\circ }{\mathbf{F}}(\frac{\partial }{\partial y^{i}})=\frac{a}{%
\left \Vert y\right \Vert }\frac{\delta }{\delta x^{i}},(i=\overline{1,n})
\label{1.7}
\end{equation}

\begin{equation}
\overset{\circ }{\mathbf{F}}^{\ast }\left( dx^{i}\text{ }\right) =-\frac{%
\left \Vert y\right \Vert }{a}\delta y^{i}\ ,\  \  \  \overset{\circ }{\mathbf{F%
}}^{\ast }\left( \delta y^{i}\  \right) =\frac{a}{\left \Vert y\right \Vert }%
dx^{i}.  \label{1.8}
\end{equation}

\section{Euler-Lagrange Equations}

Here, we obtain Euler-Lagrange equations for quantum and classical mechanics
on the almost K\"{a}hler model $\mathring{H}^{2n}=(\widetilde{TM};\overset{%
\circ }{G},\overset{\circ }{\mathbf{F}})$ of the Finsler space $F^{n}.$

Firstly, let $\overset{\circ }{\mathbf{F}}$ take a complex structure on the
almost K\"{a}hler model $\mathring{H}^{2n}$ of the Finsler space $F^{n},$
and $\left \{ x^{i},y^{i}\right \} $ be its coordinate functions.

Let semispray be the vector field $\xi $ determined by 
\begin{equation}
\xi =X^{i}\frac{\delta }{\delta x^{i}}+Y^{i}\frac{\partial }{\partial y^{i}}%
,\  \,  \label{3.1}
\end{equation}%
where $X^{i}=y^{i}=\dot{x}^{i},\ Y^{i}=\dot{y}^{i}=\ddot{x}^{i}$ and the dot
indicates the derivative with respect to time $t$.

The vector field defined by 
\begin{equation*}
V=\overset{\circ }{\mathbf{F}}(\xi )=-\frac{\left \Vert y\right \Vert }{a}%
X^{i}\frac{\partial }{\partial y^{i}}+\frac{a}{\left \Vert y\right \Vert }%
Y^{i}\frac{\delta }{\delta x^{i}}
\end{equation*}%
is named \textit{Liouville vector field} on the almost K\"{a}hler model $%
\mathring{H}^{2n}$ of the Finsler space $F^{n}.$

The maps given by $T,P:M\rightarrow R$ such that $T=\frac{1}{2}m_{i}((\dot{x}%
^{i})^{2}+(\dot{y}^{i})^{2}),P=m_{i}gh$ are said to be \textit{the kinetic
energy} and \textit{the potential energy of the system,} respectively.%
\textit{\ }Here\textit{\ }$m_{i},g$ and $h$ stand for mass of a mechanical
system having $m$ particles, the gravity acceleration and distance to the
origin of a mechanical system on the almost K\"{a}hler model $\mathring{H}%
^{2n}$ of the Finsler space $F^{n}$, respectively.

Then $L:M\rightarrow R$ is a map that satisfies the conditions; i) $L=T-P$
is a \textit{Lagrangian function, ii)} the function determined by $%
E_{L}=V(L)-L,$ is\textit{\ energy function}.

The function $i_{\overset{\circ }{\mathbf{F}}}$ induced by $\overset{\circ }{%
\mathbf{F}}$ and denoted by 
\begin{equation*}
i_{\overset{\circ }{\mathbf{F}}}\omega
(X_{1},X_{2},...,X_{r})=\sum_{i=1}^{r}\omega (X_{1},...,\overset{\circ }{%
\mathbf{F}}X_{i},...,X_{r}),
\end{equation*}%
is called \textit{vertical derivation, }where $\omega \in \wedge ^{r}{}M,$ $%
X_{i}\in \chi (M).$

The \textit{vertical differentiation} $d_{\overset{\circ }{\mathbf{F}}}$ is
given by 
\begin{equation*}
d_{\overset{\circ }{\mathbf{F}}}=[i_{\overset{\circ }{\mathbf{F}}},d]=i_{%
\overset{\circ }{\mathbf{F}}}d-di_{\overset{\circ }{\mathbf{F}}}
\end{equation*}%
where $d$ is the usual exterior derivation.

For $\overset{\circ }{\mathbf{F}}$ , the closed K\"{a}hler form is the
closed 2-form given by $\Phi _{L}=-dd_{_{\overset{\circ }{\mathbf{F}}}}L$
such that 
\begin{equation*}
d_{\overset{\circ }{\mathbf{F}}}=-\frac{\left \Vert y\right \Vert }{a}\frac{%
\partial }{\partial y^{i}}\ dx^{i}+\frac{a}{\left \Vert y\right \Vert }\frac{%
\delta }{\delta x^{i}}\  \delta y^{i}:\mathcal{F}(\widetilde{TM})\rightarrow
\wedge \widetilde{TM}.
\end{equation*}

Then we have%
\begin{eqnarray*}
\Phi _{L} &=&-dd_{\overset{\circ }{\mathbf{F}}}L=\frac{\left \Vert y\right
\Vert }{a}\frac{\delta }{\delta x^{j}}(\frac{\partial L}{\partial y^{i}}%
)dx^{j}\wedge dx^{i}-\frac{a}{\left \Vert y\right \Vert }\frac{\delta ^{2}L}{%
\delta x^{j}\delta x^{i}}dx^{j}\wedge \delta y^{i} \\
&&+\frac{\left \Vert y\right \Vert }{a}\frac{\partial ^{2}L}{\partial
y^{j}\partial y^{i}}\delta y^{j}\wedge dx^{i}-\frac{a}{\left \Vert y\right
\Vert }\frac{\partial }{\partial y^{j}}(\frac{\delta L}{\delta x^{i}})\delta
y^{j}\wedge \delta y^{i}
\end{eqnarray*}

Let $\xi $ be the second order differential equation (semispray) given by (%
\ref{3.1}). Then we calculate%
\begin{equation*}
\begin{array}{c}
i_{\xi }\Phi _{L}=\Phi _{L}\left( \xi \right) =\frac{\left \Vert y\right
\Vert }{a}\frac{\delta }{\delta x^{j}}(\frac{\partial L}{\partial y^{i}}%
)X^{i}\delta _{i}^{j}dx^{i}-\frac{\left \Vert y\right \Vert }{a}\frac{\delta 
}{\delta x^{j}}(\frac{\partial L}{\partial y^{i}})X^{i}dx^{j} \\ 
-\frac{\left \Vert y\right \Vert }{a}\frac{\partial ^{2}L}{\partial
y^{j}\partial y^{i}}X^{i}\delta y^{j}-\frac{a}{\left \Vert y\right \Vert }%
\frac{\delta ^{2}L}{\delta x^{j}\delta x^{i}}X^{i}\delta _{i}^{j}\delta
y^{i}+\frac{\left \Vert y\right \Vert }{a}\frac{\partial ^{2}L}{\partial
y^{j}\partial y^{i}}Y^{i}\delta _{i}^{j}dx^{i} \\ 
+\frac{a}{\left \Vert y\right \Vert }\frac{\delta ^{2}L}{\delta x^{j}\delta
x^{i}}Y^{i}\delta y^{j}-\frac{a}{\left \Vert y\right \Vert }\frac{\partial }{%
\partial y^{j}}(\frac{\delta L}{\delta x^{i}})Y^{i}\delta _{i}^{j}\delta
y^{i}+\frac{a}{\left \Vert y\right \Vert }\frac{\partial }{\partial y^{j}}(%
\frac{\delta L}{\delta x^{i}})Y^{i}\delta y^{j}%
\end{array}%
\end{equation*}

Energy function and its differential are 
\begin{equation*}
\begin{array}{c}
E_{L}=V(L)-L=-\frac{\left \Vert y\right \Vert }{a}X^{i}\frac{\partial L}{%
\partial y^{i}}+\frac{a}{\left \Vert y\right \Vert }Y^{i}\frac{\delta L}{%
\delta x^{i}}-L%
\end{array}%
\end{equation*}

and%
\begin{equation*}
\begin{array}{c}
dE_{L}=-\frac{\left \Vert y\right \Vert }{a}X^{i}\frac{\delta }{\delta x^{j}}%
\left( \frac{\partial L}{\partial y^{i}}\right) dx^{j}-\frac{\left \Vert
y\right \Vert }{a}X^{i}\frac{\partial ^{2}L}{\partial y^{j}\partial y^{i}}%
\delta y^{j} \\ 
+\frac{a}{\left \Vert y\right \Vert }Y^{i}\frac{\delta ^{2}L}{\delta
x^{j}\delta x^{i}}dx^{j}+\frac{a}{\left \Vert y\right \Vert }Y^{i}\frac{%
\partial }{\partial y^{j}}\left( \frac{\delta L}{\delta x^{i}}\right) \delta
y^{j}-\frac{\delta L}{\delta x^{j}}dx^{j}-\frac{\partial L}{\partial y^{j}}%
\delta y^{j}%
\end{array}%
\end{equation*}

Using $i_{\xi }\Phi _{L}=dE_{L}$, we find the expressions as follows:

\begin{equation*}
\begin{array}{c}
\frac{\left \Vert y\right \Vert }{a}[X^{i}\frac{\delta }{\delta x^{i}}+Y^{i}%
\frac{\partial }{\partial y^{i}}](\frac{\partial L}{\partial y^{i}})dx^{i}+%
\frac{\delta L}{\delta x^{i}}dx^{i} \\ 
-\frac{a}{\left \Vert y\right \Vert }[X^{i}\frac{\delta }{\delta x^{i}}+Y^{i}%
\frac{\partial }{\partial y^{i}}](\frac{\delta L}{\delta x^{i}})\delta y^{i}+%
\frac{\partial L}{\partial y^{i}}\delta y^{i}=0%
\end{array}%
\end{equation*}

and

\begin{equation*}
\frac{\left \Vert y\right \Vert }{a}\xi (\frac{\partial L}{\partial y^{i}}%
)dx^{i}+\frac{\delta L}{\delta x^{i}}dx^{i}-\frac{a}{\left \Vert y\right
\Vert }\xi (\frac{\delta L}{\delta x^{i}})\delta y^{i}+\frac{\partial L}{%
\partial y^{i}}\delta y^{i}=0
\end{equation*}

By means of integral curve $\xi (t)=\dot{\alpha}(t),$ we have the equations

\begin{equation}
\frac{\left \Vert y\right \Vert }{a}\frac{d}{dt}(\frac{\partial L}{\partial
y^{i}})+\frac{\delta L}{\delta x^{i}}=0,\frac{a}{\left \Vert y\right \Vert }%
\frac{d}{dt}(\frac{\delta L}{\delta x^{i}})-\frac{\partial L}{\partial y^{i}}%
=0  \label{3.2}
\end{equation}

such that the equations calculated in (\ref{3.2}) are named \textit{%
Euler-Lagrange equations} constructed on the almost K\"{a}hler model $%
\mathring{H}^{2n}$ of the Finsler space $F^{n}$ and thus the triple $(%
\mathring{H}^{2n},\Phi _{L},\xi )$ is called a \textit{mechanical system }on
the almost K\"{a}hler model $\mathring{H}^{2n}$ of the Finsler space $F^{n}$%
\textit{.}

\section{Hamilton Equations}

Here, we present Hamilton equations and Hamiltonian mechanical systems for
quantum and classical mechanics constructed on the almost K\"{a}hler model $%
\mathring{H}^{\ast 2n}=(\widetilde{TM};\overset{\circ }{G},\overset{\circ }{%
\mathbf{F}^{\ast }})$ of the Finsler space $F^{n}$\textit{.}

Firstly let set a 1-form

\begin{equation*}
\omega =\frac{a^{2}}{\left \Vert y\right \Vert ^{2}}x^{i}dx^{i}+y^{i}\delta
y^{i}.
\end{equation*}

Then we have the Liouville form

\begin{equation*}
\lambda =\overset{\circ }{\mathbf{F}}^{\ast }(\omega )=-\frac{a}{\left \Vert
y\right \Vert }x^{i}\delta y^{i}+\frac{a}{\left \Vert y\right \Vert }%
y^{i}dx^{i}.
\end{equation*}

and the closed form

\begin{equation*}
\phi =-d(\lambda )=\frac{a}{\left \Vert y\right \Vert }dx^{i}\wedge \delta
y^{i}
\end{equation*}

Take Hamiltonian vector field as follows: 
\begin{equation*}
X=X^{i}\frac{\partial }{\partial x^{i}}+Y^{i}\frac{\delta }{\delta y^{i}},
\end{equation*}

Then we find 
\begin{equation*}
i_{X}\Phi _{H}=\Phi _{H}(X)=\frac{a}{\left \Vert y\right \Vert }X^{i}\delta
y^{i}-\frac{a}{\left \Vert y\right \Vert }Y^{i}dx^{i}
\end{equation*}

and 
\begin{equation*}
dH=\frac{\partial H}{\partial x^{i}}dx^{i}+\frac{\delta H}{\delta y^{i}}%
\delta y^{i}.
\end{equation*}

By means of $i_{X}\Phi _{H}=dH\mathbf{,}$ the Hamiltonian vector field is
found as follows: 
\begin{equation}
X=\frac{\left \Vert y\right \Vert }{a}\frac{\delta H}{\delta y^{i}}\frac{%
\partial }{\partial x^{i}}-\frac{\left \Vert y\right \Vert }{a}\frac{%
\partial H}{\partial x^{i}}\frac{\delta }{\delta y^{i}}.  \label{4.1}
\end{equation}

Assume that a curve 
\begin{equation*}
\alpha :I\subset \mathbf{R}\rightarrow \widetilde{T^{\ast }M}
\end{equation*}%
be an integral curve of the Hamiltonian vector field $X$, i.e., 
\begin{equation}
X(\alpha (t))=\dot{\alpha}(t),\, \,t\in I.  \label{4.2}
\end{equation}%
In the local coordinates, it is obtained that 
\begin{equation*}
\alpha (t)=(x^{i},y^{i})
\end{equation*}%
and 
\begin{equation}
\overset{.}{\alpha }(t)=\frac{dx^{i}}{dt}\frac{\partial }{\partial x^{i}}+%
\frac{dy^{i}}{dt}\frac{\delta }{\delta y^{i}}.  \label{4.3}
\end{equation}

Taking (\ref{4.2}), if we equal (\ref{4.1}) and\textbf{\ }(\ref{4.3}), it
holds 
\begin{equation}
\frac{dx^{i}}{dt}=\frac{\left \Vert y\right \Vert }{a}\frac{\delta H}{\delta
y^{i}}\text{, \  \  \ }\frac{dy^{i}}{dt}=-\frac{\left \Vert y\right \Vert }{a}%
\frac{\partial H}{\partial x^{i}}  \label{4.4}
\end{equation}%
Hence, the equations introduced in (\ref{4.4}) are named \textit{Hamilton
equations} on the almost K\"{a}hler model $\mathring{H}^{\ast 2n}$\ of
Finsler manifold $F^{n}$ and then the triple $(\mathring{H}^{\ast 2n},\Phi
_{H},X)$ is said to be a \textit{Hamiltonian mechanical system }on the
almost K\"{a}hler model $\mathring{H}^{\ast 2n}$\ of Finsler manifold $F^{n}$%
\textit{.}

\section{Conclusion}

Here, we may give the following corollaries considering the above structures
and found equations by means of \cite{vasile}.

\textbf{Corollary 1}: The almost complex structure $\mathbf{F}$ $(\mathbf{F}%
^{\ast })$ on $TM$ is integrable if $(M,G)$ has constant sectional curvature 
$c$ and the function $v$ is given by $v=\frac{c-uu^{^{\prime }}}{%
2tu^{^{\prime }}-u},$ where $u>0,$ $u+2tv>0$ for all $t\geq 0.$

\textbf{Corollary 2: }Assume that $(M,G)$ has constant negative curvature $c$%
. Then $\mathring{H}^{2n}(\mathring{H}^{\ast 2n})$ with the following $u$
and $v$ is a K\"{a}hler Einstein manifold. Where $A>0,$ $t\geq 0,$ $u=A+%
\sqrt{A^{2}-2ct}>0,$ $v=\frac{1}{2t}(A-\frac{4ct}{A}-\sqrt{A^{2}-2ct})>0.$
So, the equations found in (\ref{3.2}) ((\ref{4.4})) are named \textit{%
Euler-Lagrange (Hamilton) equations } constructed on the almost K\"{a}hler
Einstein model $\mathring{H}^{2n}(\mathring{H}^{\ast 2n})$ of constant
negative curvature $c$ of the Finsler space $F^{n}$

\textbf{Corollary 3:} Assume that $(M,G)$ has constant positive curvature $c$%
. Then the tube around the zero section in $TM(T^{\ast }M)$, defined by the
condition $\left \Vert y\right \Vert ^{2}=g_{ik}y^{i}y^{k}=2t<\frac{A^{2}}{c}
$ has a structure of K\"{a}hler Einstein manifold, if the functions $u,v$
are given as follows:

$A>0,$ $u=A+\sqrt{A^{2}-2ct},$ $v=\frac{1}{2t}(A-\frac{4ct}{A}-\sqrt{%
A^{2}-2ct}),$ $0\leq t<\frac{A^{2}}{c}.$ Hence, the equations found in (\ref%
{3.2}) ((\ref{4.4})) are named \textit{Euler-Lagrange (Hamilton) equations}
constructed on the almost K\"{a}hler Einstein model $\mathring{H}^{2n}(%
\mathring{H}^{\ast 2n})$ of constant positive curvature $c$ of the Finsler
space $F^{n}.$

\textbf{Acknowledgement: }I would like to express my gratitude to Vacaru who
gave me his valuable comments and discussions to complete of this study.

\end{document}